\documentclass[11pt,conference]{IEEEtran}

\usepackage{amsmath}
\usepackage{amssymb}
\usepackage{amsfonts}
\usepackage{graphicx}
\usepackage[latin1]{inputenc}

\newcommand{\Fb}{{\bf F_2}}
\newcommand{\Fn}{\bf F_{2^n}}
\newcommand{\bigO}[1]{{\cal O}\left(#1\right)}
\newcommand{\Log}[1]{Log\left(#1\right)}
\newtheorem{Problem}{{\bf Problem}}{}
\newtheorem{Algo}{{\bf Algorithm}}{}
\newtheorem{Proposition}{{\bf Proposition}}{}

\begin{document}

\title{Finding low-weight polynomial multiples using discrete logarithm} 

\author{
\authorblockN{Frédéric Didier}
\authorblockA{INRIA Rocquencourt\\
Projet CODES,\\
Domaine de Voluceau\\
78153 le Chesnay cedex\\
Frederic.Didier@inria.fr}
\and
\authorblockN{Yann Laigle-Chapuy}
\authorblockA{INRIA Rocquencourt\\
Projet CODES,\\
Domaine de Voluceau\\
78153 le Chesnay cedex\\
Yann.Laigle-Chapuy@inria.fr
}
}

\maketitle
\begin{abstract}
Finding  low-weight multiples  of a  binary polynomial  is a  difficult problem
arising  in  the  context  of  stream  ciphers  cryptanalysis.  
The best algorithms to solve this problem are based on a time memory trade-off.  
Staying  in  this category,  we  will present  a  new  approach using  discrete
logarithm rather than a direct  representation of the  involved polynomials.   
This provides an alternative to  the previously known algorithms which improves in
some case the computational complexity. 
\footnote{This work is partially funded by CELAR/DGA.}

\end{abstract}

\section{Introduction}

Correlation  and  fast correlation  attacks  are  probably  the most  important
classes  of attacks  against  stream  ciphers based  on  linear feedback  shift
registers   (LFSRs).   They    were   originally   proposed   by   Siegenthaler
\cite{Siegenthaler85}    and     improved    by    Meier     and    Staffelbach
\cite{MeierStaffelbach}. Since then, many different versions have been proposed
\cite{CanteautTrabbia00a,JJ99a,JJ99b,JJ00}, either  very general or  adapted to
specific designs.

The basic idea is  to consider that the output of the  stream cipher is a noisy
version of  a sequence generated  by an LFSR  with the same initial  state. The
attack can  be seen as an  error-correction problem: recover  the sequence, and
therefore the initial state of the LFSR. 
To  do this  most  of the  attacks  take advantage  of  parity check  equations
existing in the sequence we are trying to recover. Those parity check equations
are in fact given by the multiples  of the feedback polynomial, and to keep the
bias as low as possible, low-weight multiples are necessary. 
As a  precomputation step, we  thus have to  find those parity  check equations
before using them in the active part of the attack.

Depending on  our objectives (finding  one or many  such multiples) and  on the
parameters (degree  of the feedback  polynomial and of the  multiples, expected
weight),  there exists different  algorithms to  find low-weight  multiples (see
\cite{CJM02,FiniaszVaudenay06}).  
We  will complete  them by  another  approach based  on the  use of  discrete
logarithm  over finite  fields.  This will  lead  to a  new  algorithm for  the
computation  of polynomials  multiples  that has  better  performance for  some
problems. 
Remark that  the complexity  of the best  method is  often still very  high for
parameters    used   in    real    cryptosystem.    Notice    also   that    in
\cite{Penzhorn_Kuhn95}  discrete  logarithms   were  already  used  to  compute
multiples of weight  $3$ and $4$.  We have generalized  this idea and improved
the complexity analysis.

The paper is organized as follows. 
Section \ref{sec:prelim} introduces some notations.  
The usual approach used to compute low-weight multiples is presented in Section
\ref{sec:usual}.  
In  Section  \ref{sec:DL},  we  detail  our  main  algorithm  and  compare  its
complexity with the algorithm of \cite{CJM02}.  
Then, we  will see  in Section \ref{sec:some}  how the complexity  is modified
when we only want to find a few multiples and not all. 
Finally,  we  will  discuss   in  Section  \ref{sec:practical}  some  important
practical points and give some experimental results in Section \ref{sec:expe}.
\section{Preliminary}
\label{sec:prelim}
\subsection{Notations}
 The problem we will be dealing with is the following.
 \begin{Problem}[Low-weight polynomial multiple]~
 \label{pb:all}

 {\bf Input:} A binary primitive polynomial $P\in\Fb[X]$ of degree $n$, and two integers $w$ and $D$.

 {\bf Output:} \emph{All} the multiples of $P$ of weight at most $w$ and degree
 at most $D$.  
 \end{Problem}

 The number of  expected such multiples of $P$  is heuristically approximated by
 $\frac{D^{w-1}}{\left(w-1\right)!2^n}$, considering that for $D$ large enough,
 the values of the polynomials of weight $w$ and degree at most $D$ are
 uniformly distributed.   Most of the time,  the degree $D$ and  the weight are
 chosen high  enough for many solutions to  exist as we need  many parity check
 equations to mount an attack. 

 It's also worth noticing  that we almost never need all the multiples.
 In fact, to mount a successful attack, one only have to find a fixed number of
 parity check equations. It is thus sufficient to find many --- but not all ---
 multiples, which might be much easier, especially if the constraint on the degree
 and the  weight are high enough.  We therefore introduce  a slightly different
 problem.

 \begin{Problem}~
 \label{pb:many}

 {\bf Input:}  A binary  primitive polynomial $P\in\Fb[X]$  of degree  $n$, and
 three integers $w$, $D$ and $B$.

 {\bf Output:} $B$ multiples of $P$ of weight at most $w$ and degree at most
 $D$, or as much as possible if there are not $B$ such multiples. 
 \end{Problem}

 \section{The classical approach}
 \label{sec:usual}
\subsection{The algorithm}

 The main idea is to use a time-memory trade-off (TMTO). 
 Set $w=q_1+q_2+1$ with $q_1\leq q_2$. 
 \begin{Algo}[TMTO]~
 \begin{itemize}
 \item             For              all             the             $q_1$-tuples
   $\Gamma=\left(\gamma_1,\ldots,\gamma_{q_1}\right)$  with   $0  <  \gamma_1  <
   \cdots <\gamma_{q_1}\leq D$, compute  and store the pairs $\left<X^{\gamma_1}
   + \cdots + X^{\gamma_{q_1}} \mod{P};\Gamma\right>$. 
 \item For  all $q_2$-tuples $\Delta =\left(\delta_1,\ldots,\delta_{q_2}\right)$
   with      $0<\delta_1<     \cdots      <\delta_{q_2}\leq      D$,     compute
   $X^{\delta_1}+\cdots+X^{\delta_{q_2}} \mod{P}$. Look in the 
 table for an  element XORing to $1$  (this can be efficiently done  by using an
 hash table). 

 If it exists, this gives
 \[ 1+\sum_{\gamma\in \Gamma}X^\gamma+\sum_{\delta\in \Delta}X^\delta = 0 \mod{P}. \]
 \end{itemize}
 \end{Algo}

 \subsection{Complexity}

The usual time-memory  trade-off is $q_1=\left\lfloor\frac{w-1}{2}\right\rfloor$
and  $q_2=\left\lceil\frac{w-1}{2}\right\rceil$,   in  order  to   balance  the
complexity of the two phases of the algorithm. 
The most time consuming part depends on the parity of $w$, as we do not have to
compute anything to find the collisions  if $q_1=q_2$. 

The memory complexity is then  $\bigO{D^{q_1}}$ (for the first phase) while the time  complexity is  $\bigO{D^{q_2}}$. 
Remark that in \cite{CJM02} the memory usage of the algorithm has been improved in order to use only $\bigO{D^{\lceil \frac{w-1}{4} \rceil}}$ bits.

\section{Using discrete logarithm}
\label{sec:DL}
\subsection{The algorithm}
In   this   section,   we   will   consider  the   field   $\Fn$   defined   as
$\Fb[x]/\left<P\right>$. 
The discrete  logarithm (with base element $x$) in this  field will  be denoted by $Log$.

Set $w=q_1+q_2+2$ with $q_1\leq q_2$.
Take  two  tuples
\[
\Gamma=\left(\gamma_1,\ldots,\gamma_{q_1}\right)\mbox{ with }0
 <\gamma_1    <   \cdots    <    \gamma_{q_1}   \leq    D
\]
and
\[
\Delta    = \left(\delta_1,\ldots,\delta_{q_2}\right)\mbox{ with }0     <     \delta_1<
 \cdots<\delta_{q_2}\leq D.
\]

Denoting    by     $L_\Gamma$    and    $L_\Delta$     the    logarithms    of
$1+\sum_{\gamma\in\Gamma}x^\gamma$          and          $1+\sum_{\delta\in
 \Delta}x^\delta$ respectively, the following equalities hold  in $\Fb[x]/\left<P\right>$:
\[
1                +               \sum_{\gamma\in\Gamma}x^\gamma               =
x^{L_\Gamma-L_\Delta}\left(1+\sum_{\delta\in\Delta}x^\delta\right) \mbox{\quad and }
\]
\[
 x^{L_\Delta-L_\Gamma}\left(1 + \sum_{\gamma\in\Gamma}x^\gamma \right) = 1
 + \sum_{\delta\in\Delta}x^\delta.
\]
Now  let $e  \in ~]-2^{n-1},2^{n-1}]$  such  that $e$  is equal  to ${L_\Gamma  -
L_\Delta}$ modulo $2^n-1$. If $e>0$, then the polynomial 
 \begin{equation}
 \label{mul1}
 \left(1        +         \sum_{\gamma\in\Gamma}x^\gamma        \right)        +
 x^e\left(1+\sum_{\delta\in\Delta}x^\delta\right)
 \end{equation}
is a multiple of $P$ with degree $\max(\gamma_{q_1},\delta_{q_2}+e)$.
If $e<0$, then the polynomial
 \begin{equation}
 \label{mul2}
 x^{-e}\left(1 + \sum_{\gamma\in\Gamma}x^\gamma \right) + \left(1
 + \sum_{\delta\in\Delta}x^\delta \right) 
 \end{equation}
is a multiple of $P$ with degree $\max(\gamma_{q_1}-e,\delta_{q_2})$.
So, if one of the two following conditions is satisfied
\begin{eqnarray*}
\label{eq:cond1} e>0 & \mbox{ and }& \delta_{q_2} + e\leq D\\
\label{eq:cond2} e<0 & \mbox{ and }& \gamma_{q_1} - e \leq D
 \end{eqnarray*}
we get a multiple of $P$ with degree at most $D$ and weight at most $w$.
We can rewrite both conditions in a single inequality
\begin{equation}
\label{cond} \gamma_{q_1} - D \le e \le D - \delta_{q_2}.   
\end{equation}
The algorithm is then straightforward. 

 \begin{Algo}[LogTMTO]~
 \begin{itemize}
 \item For  all  the $q_1$-tuples  $\Gamma=\left(\gamma_1,\ldots,\gamma_{q_1}\right)$
   with $0 < \gamma_1 < \cdots <\gamma_{q_1}\leq D$, compute
 \[L_\Gamma = \Log{1+x^{\gamma_1}+\cdots+x^{\gamma_{q_1}}}\]
 and store the pairs
 $
 \left<L_\Gamma;\Gamma\right>.
 $
 \item For all $q_2$-tuples $\Delta =\left(\delta_1,\ldots,\delta_{q_2}\right)$ with $0<\delta_1< \cdots
 <\delta_{q_2}\leq D$ compute the logarithm
 \[
 L_\Delta =\Log{1+x^{\delta_1}+\cdots+x^{\delta_{q_2}}}
 \]
  and look in
 the table for all the elements with a logarithm $L_\Gamma$ satisfying
 (\ref{cond}).
 For each of them we obtain a multiple of $P$ given by (\ref{mul1}) or (\ref{mul2}) depending on the sign of $e$.
 \end{itemize}
 \end{Algo}

Of course,  since we can decompose  all polynomials of weight  $w$ in $w-1\choose
q_1$ way, we obtain each multiple many times.

 \subsection{Complexity}

 In order to perform the second  phase, one could sort the table with increasing
 logarithms, but using an appropriate data structure like an hash table indexed
 by the  most significants bits  of the logarithm  is a lot more  efficient. As
 long as $D<2^{n/2}$, the search cost is $\bigO{1}$. 

 Once again,  we choose the parameters of  the time-memory trade-off  in order to
 balance      the     complexity     of      the     two      phases,     taking
 $q_1=\left\lfloor\frac{w-2}{2}\right\rfloor$                                 and
 $q_2=\left\lceil\frac{w-2}{2}\right\rceil$.

 As for  the classical algorithm,  the most time  consuming part depends  on the
 parity of $w$ as we do not  have to compute any logarithm in the second phase
 if $q_1=q_2$.

 The  memory usage  is  then  $\bigO{D^{q_1}}$, while  the  time complexity  is
 $\bigO{D^{q_2}}$ logarithm computations. We will see in Section \ref{ssec:log}
 that  the logarithm  can  be  computed quite  efficiently.  Actually for  many
 practical values of $n$ we can  even compute it in $\bigO{1}$. Hence we neglect
 it in Table \ref{tab:comp}.

 \begin{table}[ht]
 \caption{Comparison between TMTO and LogTMTO}
 \label{tab:comp}
 \begin{center}
 \begin{tabular}{rc|c|c|c|}
 \cline{2-5}
 & \multicolumn{2}{|c|}{$w = 2p$} & \multicolumn{2}{c|}{$w= 2p+1$} \\
 \hline
 \multicolumn{1}{|r|}{Algorithm} & Time & Memory & Time & Memory \\
 \hline
 \hline
 \multicolumn{1}{|r|}{TMTO} & $D^p$ & $D^{\lceil p/2 \rceil}$ & $D^p$ & $D^{\lceil p/2 \rceil}$ \\
 \multicolumn{1}{|r|}{LogTMTO} & $D^{p-1}$ & $D^{p-1}$ & $D^p$ & $D^{p-1}$\\
 \hline
 \end{tabular}
 \end{center}
 \end{table}

As we  can see  in Table  \ref{tab:comp}, if $w$  is even  we can improve  the time
complexity compared to the classical approach.
Heuristically, the  improvement by a factor  $D$ can be explained  by the fact 
that we  look for values in an  interval of size roughly $D$ instead of exact 
collisions.

Regarding the memory however, as explained in \cite{CJM02} the computation behind the classical algorithm can be done using only $\bigO{D^{\lceil p/2 \rceil}}$ bits.
So the discrete logarithms approach is always worse for odd $w$ and will only be of practical interest when we are looking for all the multiples of weight $4$ and maybe $6$.
After that, the memory usage just become too important.

However, we will see in the next section that when we are only looking for a small fraction of all the multiples of degree up to $D$, the discrete logarithms method can be quite efficient. 

\section{Find many but not all}
 \label{sec:some}

We deal in this section with the problem of finding a small proportion of all the multiples of weight $w$ and degree at most $D$ (Problem \ref{pb:many}).
If the number $B$ of polynomials we want is small enough, depending on the parameters, we can do better than the previous algorithms.  

A very basic approach is to try random polynomials of weight $w$ until we actually find a multiple. 
In expectation we will then find a multiple every $2^n$ polynomials tried.
We can also do the same using discrete logarithms.
By computing logarithms for polynomials $A$ of weight $w-1$ and degree less than $D$, 
we can obtain  easily low-weight multiples of type $A+x^{\Log{A}}$ if the logarithm is at most $D$. 
The expectation here is to find a multiple every $2^n/D$ iterations and we have won a factor $D$. 

However, the best methods to solve this problem are once again TMTO.
The algorithms are just simple variations of the previous ones when we put the elements in the hash table one
by one and stop when we have found enough multiples.

Applying the birthday paradox, we can thus find with the basic algorithm a multiple with a time and memory 
complexity of $\bigO{\sqrt{2^n}}$ in average.
Using discrete  logarithms, we will find  a multiple as soon  as two logarithms
have a distance by approximately $D$. 
The complexity is then in $\bigO{\sqrt{\frac{2^n}{D}}}$ both in time and memory.
Remark that in this case one cannot use the improvement of \cite{CJM02} to gain memory.
There is also  another approach based on Wagner's  generalized birthday paradox
(see \cite{wagner,FiniaszVaudenay06}) that can be usefull when $w$ is large.
Its complexity is in $\bigO{2^a2^{n/(a+1)}}$ for a $a$ such that $\binom{D}{(w-1)/2^a} \ge 2^{n/(a+1)}$.

As a conclusion to this section, when computing logarithms in $\Fn^\star$ is easy, we can
gain a factor $\sqrt{D}$ in time and memory to find a multiple.
Notice also that in practice when we need many multiples, we can design an algorithm between 
the one that compute all the multiples and the one presented here in order to get the best performance.
We will see an illustration of this in Section \ref{sec:expe}.

\section{Practical Considerations}
\label{sec:practical}

 \subsection{Bounds on the degree}
 First of all, it is worth noticing that it is not necessary to compute all the multiples up to the degree $D$ to take all $q_2$-tuples up to the degree $D$.

 As a polynomial of weight $w$ has many representations as a sum of a polynomial of weight
 $q_1+1$ and $q_2+1$ respectively, we can choose the one with the smallest $q_2$-tuple.

 \begin{Proposition}
 Let $M=1+\sum_{i\in I} X^i$ be a multiple of $P$ of weight $w=q_1+q_2+2$ and degree
 at most $D$. 

 Then there exists  an integer $1\leq e\leq D$ and two  polynomials $A$ and $B$
 of respective weight $q_1$ and $q_2$ and of degree respectively at most $D$ and
 at most $\frac{Dq_2}{w-1}$ such that
 $M=\left(1+A\right)+X^e\left(1+B\right)
 \mbox{ or }
 X^e\left(1+A\right)+\left(1+B\right).$
 \end{Proposition}

 With  the usual  trade-off, we  can restrict  ourselves to  the  degree $D/2$,
 dividing the cost of the second phase approximately by a factor $2^{w/2}$.

\subsection{How to compute logarithms}
\label{ssec:log}
In  practice, it  is important  to  compute efficiently  discrete logarithms  in
$\Fn^\star$ and hopefully there exists well studied algorithms to do that.  
It  is important  to  take  into account  that  we are  going  to compute  many
logarithms  and not  only  one.   All the  efficient  algorithms for  computing
logarithms      (Baby-step       Giant-step,      Pohlig-Hellman      algorithm
\cite{pohlig-hellman-dlog}  and Coppersmith  algorithm  \cite{Copp1,Copp2}) can
 profit  from a bigger  precomputation step that  can be done once  and for
all. 
For instance, if  $2^n-1$ is smooth enough, one can  tabulate the logarithms in
all  the subgroups  of $\Fn^\star$  to make  the Pohlig-Hellman  algorithm very
efficient. 
In this case, a subsequent discrete logarithm computation can be done in $\bigO{1}$.
This   approach  can   be   used  for   all   the  $n$   up   to  $78$   except
$\{37,41,49,59,61,62,65,67,69,71,74,77\}$. In addition we have listed in Table
\ref{tab:PH} some larger  $n$ for which it is  applicable and the corresponding
memory requirement. 
Notice that  a full  tabulation corresponds to  a Giant-step  of 1 and  that by
increasing a little this Giant-step, we can efficiently deal with more values of $n$
.

\begin{table}[ht]
\caption{Memory usage for a fully tabulated Pohlig-Hellman algorithm and some smooth $2^n-1$}
\label{tab:PH}
\begin{center}
\begin{tabular}{|c||ccccc|}
\hline
$n$&$53$&$96$&$110$&$156$&$210$\\
\hline
memory&$439$MB&$510$MB&$1.7$GB&$940$MB&$201$MB\\
\hline
\end{tabular}
\end{center}
\end{table}

This leads to a very easy and efficient implementation as we will see in Section \ref{sec:expe}. 
Moreover, for the most useful cases (that is $w\in\left\{3,4,5\right\}$) we have to
compute logarithms of the form $\Log{1+x^i}$.
This logarithm is know as the Zech's  logarithm of $i$, and we can exploit some
properties  of   Zech's  logarithm  (see   \cite{Huber90})  to  speed   up  the
computation. 
Actually, by computing one Zech logarithm we get $6 n$ other logarithms for free.
Of course not all  of them are useful for us, but  the computation time can be
divided by a factor of at least $2$. 

\section{Experimental result}
\label{sec:expe}

We have implemented our algorithm in C to test its efficiency.
The computer used for our experiments is a $3.6$GHz Pentium$4$ with $2$MB of cache and $2$GB of RAM. 

\subsection{Problem 1}

We give in Table \ref{tab:efficace} the timings to find all the multiples of weight $w$ up to degree $D$ of the polynomial

$
P=x^{53} + x^{47} + x^{45} + x^{44} + x^{42} + x^{40} + x^{39} +
  x^{38} + x^{36} + x^{33} + x^{32} + x^{31} + x^{30} + x^{28} +
  x^{27} + x^{26} + x^{25} + x^{21} + x^{20} + x^{17} + x^{16} +
  x^{15} + x^{13} + x^{11} + x^{10} + x^7 + x^6 + x^3 + x^2 + x^1 + 1.
$

As explained in the previous section, we used a fully tabulated Pohlig-Hellman.

\begin{table}[ht]
\caption{Problem $1$: find all the multiples up to degree $D$}
\label{tab:efficace}
\[
\begin{array}{|c||c|c|c||c|c|c|}
\hline
n&\multicolumn{6}{c|}{53}\\
\hline
w&\multicolumn{3}{c||}{4}&\multicolumn{3}{c|}{5}\\
\hline
\log_2(D) & 20   & 22       & 28         & 13         & 14          & 16\\
\hline
\mbox{time}      & 47'' & 2' 02'' & 1 h 52' & 4' 11'' & 14' 40'' & 3 h 33'\\
\hline
\end{array}
\]
\end{table}

We can see  that the algorithm is, as expected, very efficient for weight $4$ as its complexity is
linear in the degree  $D$, both for time and memory (to be compared to a
quadratic complexity for the classical approach). 

We were also able to compute all the multiples of weight $5$ and degree up to $2^{16}$
of a polynomial of degree $53$ within a few hours. 
But for the degree $5$ the algorithm of \cite{CJM02} is more efficient.

\subsection{Problem 2}

\begin{figure}[ht]
\begin{picture}(0,0)%
\includegraphics[scale=0.7]{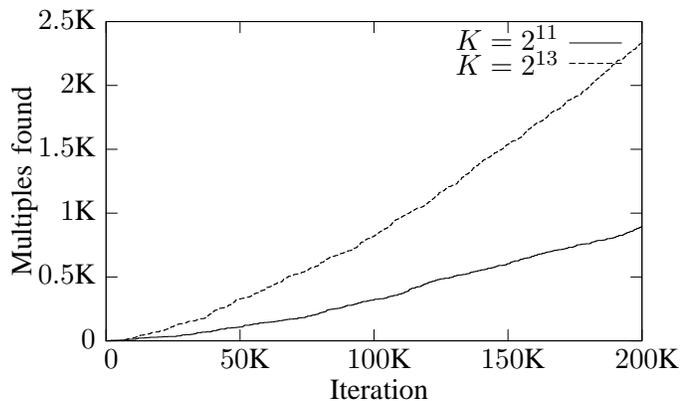}%
\end{picture}%
\begingroup
\setlength{\unitlength}{0.0140bp}%
\begin{picture}(18000,10800)(0,0)%
\put(2475,1650){\makebox(0,0)[r]{\strut{} $0$}}%
\put(2475,3370){\makebox(0,0)[r]{\strut{} $0.5$K}}%
\put(2475,5090){\makebox(0,0)[r]{\strut{} $1$K}}%
\put(2475,6810){\makebox(0,0)[r]{\strut{} $1.5$K}}%
\put(2475,8530){\makebox(0,0)[r]{\strut{} $2$K}}%
\put(2475,10250){\makebox(0,0)[r]{\strut{} $2.5$K}}%
\put(2750,1100){\makebox(0,0){\strut{} $0$}}%
\put(6356,1100){\makebox(0,0){\strut{} $50$K}}%
\put(9962,1100){\makebox(0,0){\strut{} $100$K}}%
\put(13569,1100){\makebox(0,0){\strut{} $150$K}}%
\put(17175,1100){\makebox(0,0){\strut{} $200$K}}%
\put(550,5950){\rotatebox{90}{\makebox(0,0){\strut{}Multiples found}}}%
\put(9962,275){\makebox(0,0){\strut{} Iteration}}%
\put(14950,9675){\makebox(0,0)[r]{\strut{}$K=2^{11}$}}%
\put(14950,9000){\makebox(0,0)[r]{\strut{}$K=2^{13}$}}%
\end{picture}%
\endgroup
\caption{Evolution of  the number of multiples  of weight $7$  and degree lower
  than $2^{15}$ found with precomputed logarithms up to degree $K$}
\label{fig:pb2}
\end{figure}

With the same polynomial of degree $n=53$, we also looked for multiples with an
higher weight $w=7$, and degree at most $D=2^{15}$.  
In order to do that, we precomputed all  the trinomials ($1+x^{\gamma_1}+x^{\gamma_2}$) up  to the  degree $K$,  which corresponds  to $q_1=2$, instead of $3$ for the optimal trade-off. 
We then computed  many discrete logarithm of random polynomials ($1+x^{\delta_1}+x^{\delta_2}+x^{\delta_3} + x^{\delta_4}$) in order to find multiples of weight $7$.
The results are given in Figure \ref{fig:pb2} where we see that a bigger precomputation can greatly
improve the performance.

\section{Conclusion}

In this  paper, we devised  an algorithm  to find low-weight multiples  of a
given binary polynomial that appears to be efficient for two cases that actually
occur in practice.

The first case is when we are looking for all the multiples of weight $4$ and degree at most $D$ of a given polynomial of degree $n$. 
The complexity is then in $\bigO{D}$ discrete logarithms computation in $\Fn^\star$ where the other approach run in $\bigO{D^2}$.
So the  best algorithm will depends  on the complexity of  a discrete logarithm
computation in $\Fn^\star$ which can be smaller than $D$ in many practical situations. 
Notice that our algorithm may also give better performance for multiples of weight $6$.

The other case where discrete logarithms can be useful is when we are only looking
for a small fraction of all the possible multiples. 
The complexity to find one of them is then $\bigO{\sqrt{\frac{2^n}{D}}}$ logarithm computations.

\section*{Acknowledgment}
The authors would like to thank Anne Canteaut and Jean-Pierre Tillich for their
helpful insights on the subject.

\end{document}